\documentclass[11pt]{pazh_engl}

\usepackage[T2A]{fontenc}
\usepackage[utf8]{inputenc}

\usepackage{latexsym}
\usepackage{amssymb}

\usepackage{graphics}
\usepackage{graphicx}
\usepackage{latexsym}
\usepackage{amssymb}
\usepackage{amsmath}
\usepackage{hyperref}
\usepackage[hyphenbreaks]{breakurl}

\begin{document}

\sloppypar

\title{\bf Edge-on galaxies in the Hubble Ultra Deep Field}

\author{V.P. Reshetnikov\inst{1,2}, P.A. Usachev\inst{1,2}, 
S.S. Savchenko\inst{1,2}} 

\institute{St.Petersburg State University, Universitetskii pr. 28, 
St.Petersburg, 198504 Russia
\and
Special Astrophysical Observatory, Russian Academy of Sciences, 
Nizhnii Arkhyz, 369167 Russia
}


\abstract{We studied a sample of 58 edge-on spiral galaxies 
at redshifts $z \sim 1$ selected in the Hubble Ultra Deep Field. For all 
galaxies we analyzed the 2D brightness distributions in the $V_{606}$
and $i_{775}$ filters and measured the radial ($h_r$) and vertical ($h_z$) 
exponential scales of the brightness distribution. We obtained 
evidence that the relative thickness of the disks of distant galaxies, i.e., 
the ratio of the vertical scale height and radial scale length ($h_z/h_r$), 
on average, exceeds the relative thickness of the disks of nearby spiral 
galaxies. The vertical scale height $h_z$ of the stellar disks of galaxies 
shows no big changes at $z \leq 1$. The possibility of the evolution of the 
radial scale length $h_r$ for the brightness distribution with redshift is
discussed.
}

\titlerunning{Edge-on galaxies in the HUDF}

\maketitle

\section{Introduction}

A study of edge-on spiral galaxies allows a number of important problems 
of extragalactic astronomy to be investigated: the structure and stability 
of galactic disks, the properties and distribution of dust in
them, the contribution of dark matter to the structure of galaxies, 
the large-scale distribution of galaxies, etc. (see, e.g., van der Kriut 
and Searl 1981; Zasov et al. 1991; de Grijs 1998; Mosenkov et al. 2010,
2016; Bizyaev et al. 2017; Makarov et al. 2018; and references therein). 
The preceding papers were devoted mostly to edge-on galaxies in the local
Universe. Only in a few papers the vertical structure
of distant objects was studied. For example, Reshetnikov et al. (2003) 
investigated edge-on galaxies in the Hubble deep fields
north (HDF-N) and south (HDF-S).
They found that the relative thickness of the stellar disks
of galaxies at redshifts $z \sim 1$ exceeds the relative
thickness of the disks of nearby galaxies by a factor of 1.5–2. This 
conclusion was confirmed when analyzing the structure of galaxies in 
the Hubble Ultra Deep Field (hereafter HUDF) (Elmegreen et al. 2005;
B. Elmegreen and D. Elmegreen 2006).

The goal of our paper is a photometric study of edge-on spiral galaxies 
in the HUDF. The main differences between our paper and the previously 
published studies are: an analysis of the complete two-dimensional (2D) 
brightness distributions instead of the one-dimensional profiles, using 
the spectroscopic redshifts for most objects, and a larger size of the
sample of edge-on galaxies.

All of the numerical values in our paper are given
for the cosmological model with the Hubble constant of
of 70 km s$^{-1}$ Mpc$^{-1}$ and $\Omega_m=0.3$, $\Omega_{\Lambda}=0.7$.

\section{The sample of galaxies and data reduction}

To study the edge-on spiral galaxies, we used the HUDF frames in the F606W 
(hereafter $V_{606}$) and F775W ($i_{775}$) filters (Beckwith et al. 2006). 
In these color bands the HUDF images are deeper than those
in other original filters. The pixel size is 0.03$''$. In the first step, 
on the field image in the $V_{606}$ filter we selected 901 galaxies with an 
apparent flattening $b/a \leq 0.55$, an area $\geq$24 pixels, and 
$S/N > 3$ in each pixel using the SExtractor package (Bertin and
Arnouts 1996). Such a soft constraint on the flattening was used 
in order not to throw away the galaxies with close neighbors. In several 
cases, SExtractor does not separate them, but detects them as a
single object. Next, based on a visual examination of the images for the 
objects in different filters and with different brightness contrasts, 
we left 77 candidates for edge-on galaxies in the sample.

To analyze the photometric structure (decomposition) of the galaxies, 
we used the {\large Imfit} package (Erwin 2015) with the PSF 
(point spread function) generated for the HUDF by the Tiny Tim code (Krist
et al. 2011). A comparison of the PSF with the sample objects images showed 
that the disks of all galaxies are resolved with confidence in both 
radial and vertical directions.

In the selected candidates for edge-on galaxies the bulges are noticeable 
approximately in a quarter of the objects. In most cases, these bulges are 
faint and are on the verge of resolution. Therefore, to describe the
photometric structure of the galaxies, we chose the simplest model of an 
edge-on double exponential disk (see, e.g., van der Kriut and Searl 1981):

$$
I(r, z) = 
I_{0,0} 
  \left(\frac{r}{h_r}\right) K_1 
    \left(\frac{r}{h_r}\right) e^{-z/h_z},
$$
    
where $I_{0,0}$, $h_r$, and $h_z$ are the central surface brightness, 
radial and vertical exponential scales  of the disk, respectively, 
and $K_1$ is a modified first-order Bessel function.
The bulges, if they were visible, were fitted by a S\'ersic
function. The nearby objects projected onto the galaxies under study were 
masked before the {\large Imfit} operation. If, however, the area of
the hampering objects was too large, then during the decomposition they 
were fitted by a combination of different model functions and were subtracted.

An analysis of the residual images (the original image minus the model 
one) revealed that 19 of the 77 galaxies either are not edge-on galaxies 
or have a very complex and asymmetric structure. These objects were 
excluded from the subsequent consideration. Thus, the final sample of 
edge-on galaxies studied in our paper consists of 58 objects. Examples 
of our photometric analysis for three galaxies are given in Fig. 1.

\begin{figure*}
\centering
\includegraphics[width=0.8\textwidth, angle=0, clip=]{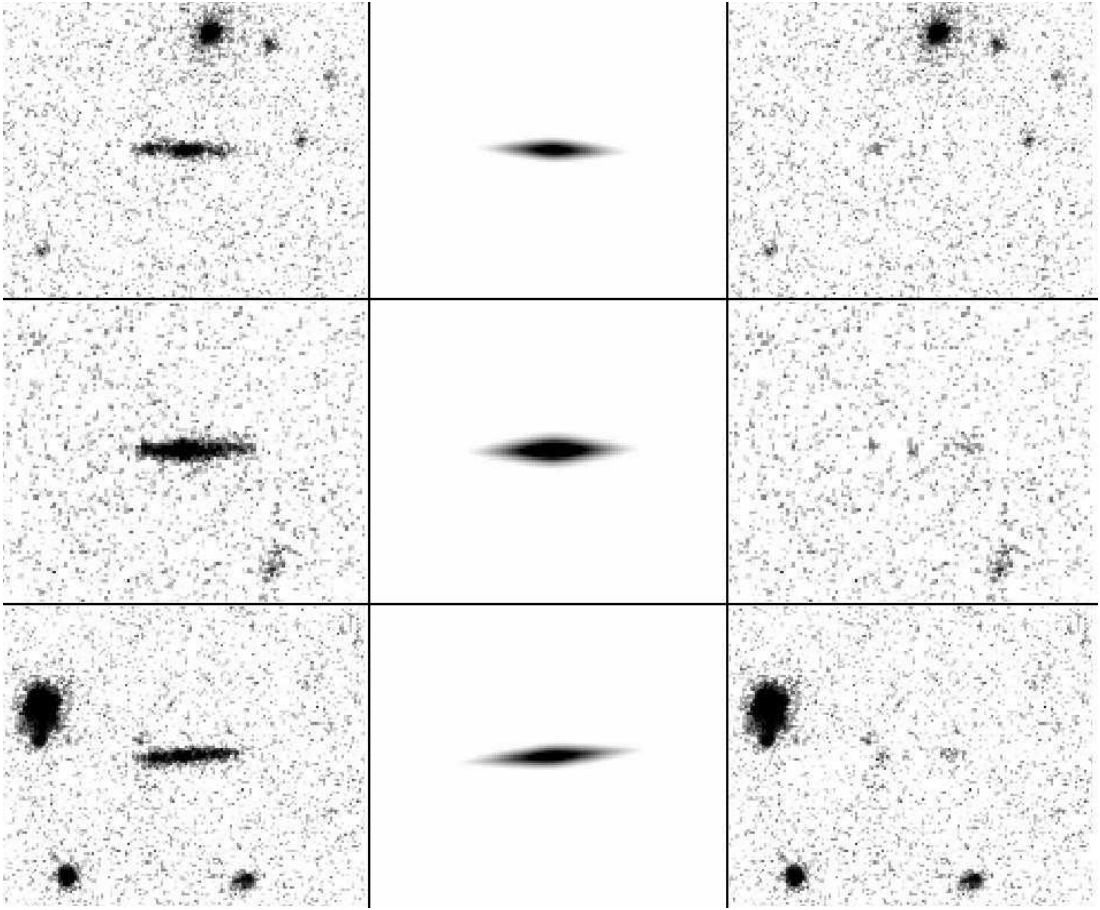}\\
\caption{Examples of photometric modeling for three sample galaxies. 
From left to right: the original image, the model, and the
difference of the original and model images. The upper, middle, and lower 
rows show, respectively, the images of galaxies N\,9 (the frame size along 
the horizontal axis is 4.6$''$), N\,22 (the corresponding size is 3.8$''$), 
and N\,27 (4.6$''$) from Table 1.}
\label{fig1}
\end{figure*}

For 33 sample galaxies we took their spectroscopic redshifts from Inami 
et al. (2017) and Rafelski et al. (2015). For 23 objects we used the 
photometric $z$ from Rafelski et al. (2015) (BPZ redshifts). For
two galaxies the redshifts were not found.

The final characteristics of the candidates for edge-on galaxies in the 
HUDF are listed in Table\,1. Column 1 in the table gives the object ordinal 
number in the sample, columns 2 and 3 provide the coordinates of the 
galactic center on the original HUDF image in pixels, the next columns 
give the galaxy number from Coe et al. (2006) and its apparent
F606W magnitude in the AB system of magnitudes (Rafelski et al. 2015). 
Column 6 gives the redshifts, with the photometric $z$ being marked by colons.
For galaxy N\,47 the photometric $z$ is very low (0.04). This value leads 
to implausible characteristics of the galaxy and, therefore, we do not 
use it in the subsequent analysis. Columns 7 and 8 in Table\,1
provide the classes introduced by us, which reflect the subjective 
probability that a galaxy belongs to edge-on ones ($eon$) and the quality 
of the photometric decomposition ($fit$) (1 means the highest probability
and quality, 3 means a low one).

The decomposition results are summarized in the last columns of the table. 
(We are interested in the scales of the brightness distribution 
$h_r$ and $h_z$, so that the central surface brightnesses of the galaxies 
are not discussed in the paper.) For both exponential scales the 
first and second numbers refer to the $V_{606}$ and $i_{775}$ filters, 
respectively. For three galaxies (N\,1, N\,15, and N\,47) the results in 
Table\,1 are presented in arcseconds. The typical measurement error of 
the scales yielded by the {\large Imfit} code is less than 10\%.


\begin{table*}
\caption{Edge-on galaxies in the HUDF}
\begin{center}
\begin{tabular}{|c|cc|c|c|c|c|c|c|c|}
\hline
N & X &  Y & CoeID & $V_{606}$ & $z$  & $eon$ & $fit$ & $h_r$ (kpc) & $h_z$ (kpc) \\
\hline               
1  & 1488 & 4578 &      &      &                   & 1 & 3 & 0.21$''$~~0.21$''$ & 0.03$''$~0.03$''$ \\
2  & 1744 & 6169 & 6870 &27.05 &             1.14  & 2 & 1 & 1.76~~1.48 & 0.40~~0.44 \\
3  & 1812 & 5234 & 4907 &26.48 & \hphantom{:}2.02: & 3 & 1 & 1.36~~1.31 & 0.43~~0.42 \\
4  & 1941 & 4739 & 3840 &26.77 & \hphantom{:}0.59: & 3 & 1 & 1.40~~1.20 & 0.37~~0.37 \\
5  & 2084 & 4477 & 3299 &25.51 &             1.22  & 2 & 2 & 2.11~~1.92 & 0.46~~0.49 \\
6  & 2148 & 6006 & 6478 &25.81 & \hphantom{:}2.94: & 2 & 2 & 2.51~~2.42 & 0.60~~0.57 \\
7  & 2697 & 6387 & 7022 &25.03 &             1.55  & 3 & 3 & 2.79~~2.50 & 0.81~~0.77 \\
8  & 2701 & 5920 & 6278 &26.62 &             0.94  & 2 & 2 & 2.77~~2.47 & 0.41~~0.42 \\
9  & 2738 & 4401 & 3315 &27.58 & \hphantom{:}0.97: & 2 & 1 & 1.99~~1.75 & 0.31~~0.34 \\
10 & 3240 & 3843 & 2332 &26.29 &             0.52  & 3 & 2 & 1.18~~1.08 & 0.39~~0.41 \\
11 & 3401 & 2811 & 1057 &24.83 & \hphantom{:}0.74: & 1 & 2 & 3.66~~3.28 & 0.55~~0.59 \\
12 & 3631 & 5784 & 5995 &25.04 &             0.95  & 3 & 2 & 1.22~~1.20 & 0.52~~0.52 \\
13 & 3973 & 6512 & 7269 &24.23 &             0.73  & 1 & 2 & 3.50~~3.51 & 0.70~~0.74 \\
14 & 4140 & 6814 & 8801 &25.78 &             1.31  & 3 & 1 & 2.37~~2.35 & 0.57~~0.60 \\
15 & 4168 & 6469 &      &      &                   & 3 & 1 & 0.18$''$~~0.16$''$ & 0.05$''$~0.04$''$ \\
16 & 4186 & 4249 & 3097 &26.56 & \hphantom{:}2.36: & 3 & 1 & 1.67~~1.59 & 0.63~~0.59 \\
17 & 4213 & 3054 & 1242 &26.34 & \hphantom{:}2.02: & 2 & 2 & 4.04~~3.72 & 0.69~~0.68 \\
18 & 4312 & 8261 & 9414 &27.25 & \hphantom{:}1.38: & 3 & 1 & 1.96~~1.66 & 0.44~~0.46 \\
19 & 4362 & 1468 & 163  &27.43 & \hphantom{:}0.65: & 3 & 1 & 1.09~~1.07 & 0.22~~0.23 \\
20 & 4462 & 2075 & 521  &25.93 & \hphantom{:}1.04: & 1 & 2 & 2.53~~2.53 & 0.48~~0.51 \\
21 & 4673 & 7460 & 8259 &27.33 &             0.68  & 3 & 1 & 1.17~~1.12 & 0.31~~0.31 \\
22 & 4675 & 5841 & 6143 &26.97 &             1.02  & 1 & 1 & 1.67~~1.45 & 0.33~~0.38 \\
23 & 4780 & 1333 & 95   &26.21 & \hphantom{:}1.76: & 2 & 2 & 1.64~~1.56 & 0.43~~0.45 \\
24 & 4833 & 4000 & 2652 &25.17 &             0.68  & 1 & 2 & 2.46~~2.40 & 0.46~~0.47 \\
25 & 4837 & 2673 & 966  &26.43 &             1.04  & 2 & 3 & 1.56~~1.44 & 0.36~~0.38 \\
26 & 4852 & 2255 & 666  &25.94 &             1.16  & 3 & 1 & 1.13~~1.02 & 0.33~~0.33 \\
27 & 4942 & 4289 & 3101 &27.14 &             1.37  & 1 & 1 & 2.56~~2.28 & 0.34~~0.39 \\
28 & 5022 & 8056 & 9171 &26.17 &             0.68  & 3 & 1 & 1.44~~1.27 & 0.37~~0.37 \\
29 & 5023 & 8224 & 9425 &26.18 & \hphantom{:}1.79: & 2 & 2 & 1.65~~1.47 & 0.37~~0.39 \\
30 & 5074 & 2004 & 446  &25.40 &             1.10  & 3 & 1 & 1.54~~1.57 & 0.36~~0.41 \\
31 & 5404 & 5620 & 5615 &25.88 &             1.10  & 3 & 1 & 2.17~~1.59 & 0.77~~0.82 \\
32 & 5560 & 7272 & 8351 &26.39 & \hphantom{:}1.75: & 3 & 2 & 1.56~~1.49 & 0.42~~0.44 \\
33 & 5597 & 9353 & 9848 &26.53 &             1.04  & 3 & 2 & 2.29~~1.96 & 0.58~~0.58 \\
34 & 5650 & 9326 & 9974 &25.12 &             1.02  & 2 & 3 & 2.46~~2.21 & 0.61~~0.62 \\
35 & 5692 & 2437 & 833  &26.88 &             1.55  & 2 & 1 & 1.63~~1.45 & 0.31~~0.33 \\
36 & 5781 & 7049 & 8624 &25.72 &             0.83  & 1 & 1 & 4.13~~3.70 & 0.56~~0.57 \\
37 & 5811 & 5988 & 6038 &24.40 &             0.67  & 1 & 2 & 4.48~~4.27 & 1.08~~1.00 \\
38 & 5959 & 3306 & 1612 &26.51 & \hphantom{:}1.76: & 3 & 1 & 1.38~~1.36 & 0.47~~0.45 \\
39 & 6124 & 4282 & 3178 &26.21 & \hphantom{:}1.91: & 3 & 2 & 4.43~~3.84 & 0.73~~0.63 \\
40 & 6178 & 8569 & 9807 &25.88 &             0.77  & 1 & 2 & 2.62~~2.45 & 0.39~~0.43 \\
41  & 6416 & 8780 & 9834 &22.55 &             0.43  & 2 & 2 & 2.69~~3.00 & 0.77~~1.03 \\
42  & 6462 & 4440 & 3418 &26.59 & \hphantom{:}3.96: & 3 & 3 & 1.58~~1.89 & 0.46~~0.49 \\
43  & 6486 & 6320 & 6922 &25.55 &             1.26  & 1 & 3 & 5.26~~5.29 & 0.55~~0.61 \\
44  & 6491 & 7924 & 9139 &25.83 &             1.84  & 3 & 3 & 1.73~~1.58 & 0.36~~0.42 \\
45  & 6746 & 7127 & 8372 &23.22 &             0.53  & 3 & 3 & 2.63~~2.41 & 0.55~~0.57 \\
46  & 6785 & 2367 & 735  &25.72 & \hphantom{:}1.12: & 3 & 2 & 1.63~~1.30 & 0.44~~0.43 \\
47  & 6789 & 5075 & 4661 &26.63 & \hphantom{:}0.04: & 2 & 1 & 0.16$''$~~0.15$''$ & 0.05$''$~0.05$''$  \\
48  & 6894 & 7813 & 7737 &24.39 &             0.53  & 3 & 2 & 1.39~~1.37 & 0.42~~0.45 \\
49  & 6976 & 2949 & 1253 &27.44 & \hphantom{:}2.88: & 3 & 1 & 1.39~~1.24 & 0.42~~0.41 \\
50  & 7079 & 5197 & 4835 &25.48 &             1.32  & 3 & 2 & 1.55~~1.44 & 0.58~~0.56 \\
\hline
\end{tabular}
\end{center}
\end{table*}

\addtocounter{table}{-1}


\begin{table*}
\caption{Edge-on galaxies in the HUDF (cont.)}
\begin{center}
\begin{tabular}{|c|cc|c|c|c|c|c|c|c|}
\hline
N & X &  Y & CoeID & $V_{606}$ & $z$  & $eon$ & $fit$ & $h_r$ (kpc) & $h_z$ (kpc) \\
\hline               
51  & 7429 & 5431 & 5408 &26.95 &             1.42  & 3 & 2 & 1.27~~1.27 & 0.38~~0.35 \\
52  & 7792 & 4286 & 3143 &25.23 &             1.10  & 3 & 2 & 2.59~~2.14 & 0.60~~0.54 \\
53  & 7856 & 3452 & 1732 &25.57 & \hphantom{:}0.66: & 3 & 2 & 1.62~~1.65 & 0.43~~0.46 \\
54  & 7905 & 3600 & 2017 &26.53 & \hphantom{:}0.63: & 2 & 1 & 1.42~~1.34 & 0.31~~0.31 \\
55  & 8020 & 4758 & 3871 &26.16 &             0.67  & 1 & 2 & 1.93~~1.75 & 0.35~~0.36 \\
56  & 8614 & 5763 & 5898 &26.01 & \hphantom{:}1.45: & 3 & 3 & 1.23~~1.24 & 0.57~~0.57 \\
57  & 8800 & 4950 & 4321 &26.24 &             1.10  & 3 & 3 & 1.86~~1.74 & 0.48~~0.48 \\
58  & 9259 & 5065 & 4360 &24.54 & \hphantom{:}0.14: & 3 & 2 & 0.82~~0.77 & 0.26~~0.27 \\
\hline
\end{tabular}
\end{center}
\end{table*}

\section{Results and discussion}

\subsection{General characteristics of the sample galaxies}

\begin{figure}
\includegraphics[width=0.7\textwidth, angle=-90, clip=]{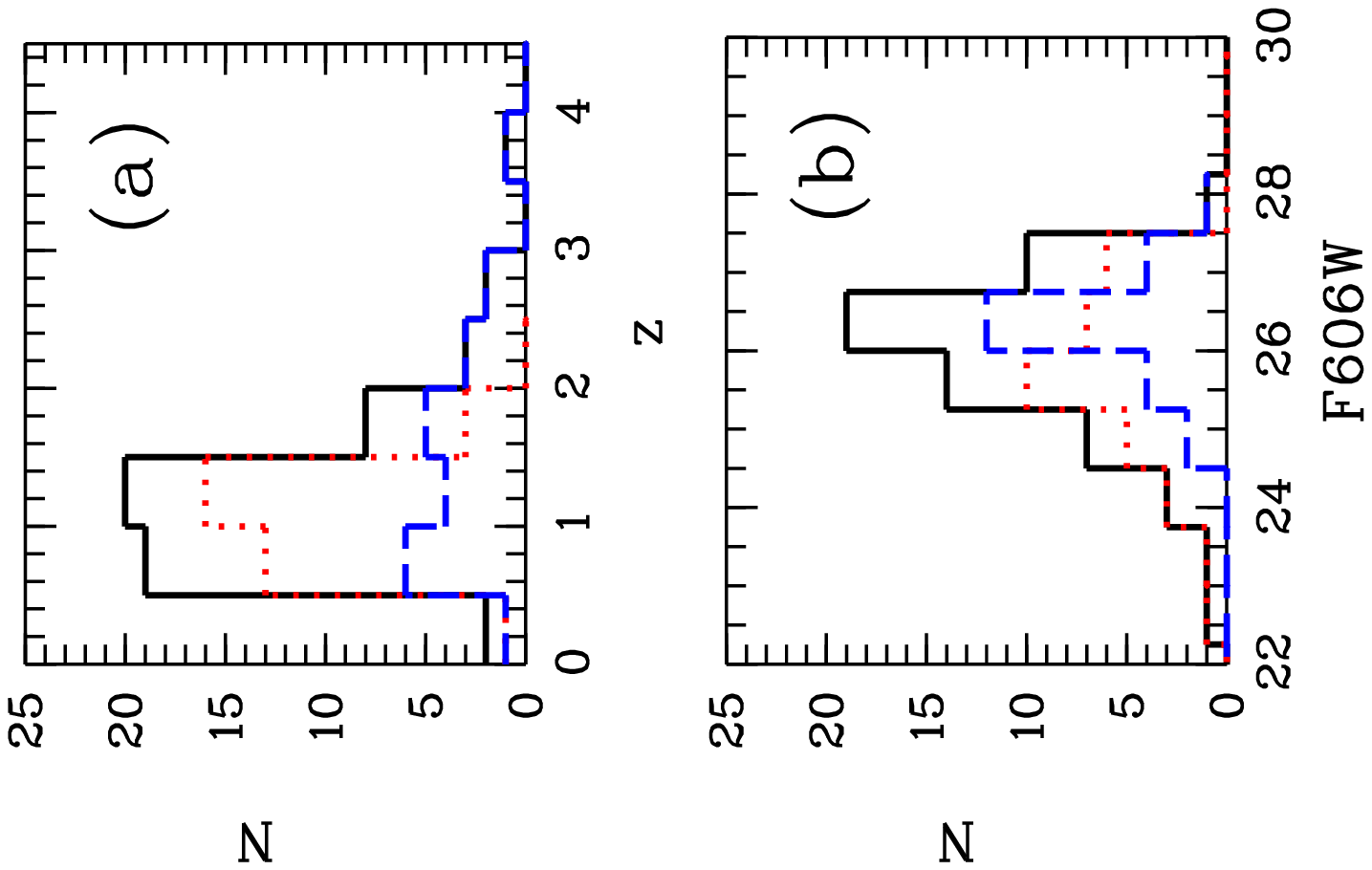}\\
\caption{(a) Distribution of the sample galaxies in redshifts (the dotted (red)
and dashed (blue) lines indicate the distribution of the galaxies with spectroscopic 
and photometric $z$, respectively; the solid line indicates the combined 
distribution); (b) the same for the apparent $V_{606}$ magnitudes.}
\label{fig2}
\end{figure}

Figure\,2 shows the distributions of the galaxies by the redshift and apparent 
$V_{606}$ magnitude. The mean redshift of the galaxies we consider is 
$\left< z \right> = 1.23 \pm 0.69$ (here and below, the unbiased sample 
variance is given as an error). The galaxies with spectroscopic $z$ are, 
on average, nearer than the objects with photometric estimates 
($\left< z \right> = 1.01 \pm 0.34$ vs. 
$\left< z \right> = 1.57 \pm 0.92$) and are brighter (Fig.\,2).

For the cosmological model adopted in this paper the redshift $z = 1.2$ 
corresponds to the time elapsed after the beginning of cosmological expansion, 
$\sim$5 Gyr. Consequently, the epochs at which we study the galaxies in 
the HUDF and the regions of the nearby Universe ($z \approx 0$) are spaced 
more than 8 Gyr apart. It is hoped that on such a long time scale we will be
able to find evidence for the evolution of the global structure of disk 
galaxies.

Figure\,3 shows the distribution of the galaxies at $z \leq 2$ in luminosity. 
The redshift constraint was used, because the photometric $z$ (only these 
are known for galaxies at $z > 2$) for distant galaxies are generally
less accurate than those for nearer ones. To find the absolute $B$ magnitudes 
($M(B)$), we used the results by Sirianni et al. (2005) and for all objects
applied the $k$-correction for an Sc galaxy from Bicker et al. (2004). 
The mean observed luminosity for the edge-on galaxies in the HUDF is 
$\left< M(B) \right> = -18.^{\text{m}}5 \pm 1.^{\text{m}}4$. Given the 
correction for internal absorption, which reaches 
$\sim 1^{\text{m}} - 1.^{\text{m}}5$ for edge-on galaxies (see, e.g., Tully et al. 
1998), the luminosities of these galaxies seen face-on, on average, reach
values in the range from --19$^{\text{m}}$ to --20$^{\text{m}}$.

\begin{figure}
\includegraphics[width=0.35\textwidth, angle=-90, clip=]{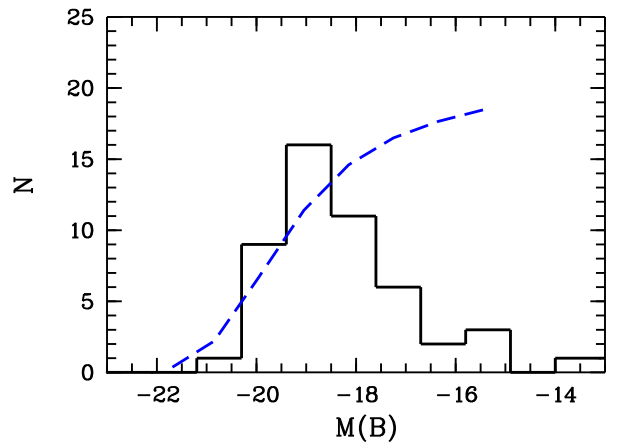}\\
\caption{Distribution of the edge-on galaxies in the HUDF at $z \leq 2$ in 
absolute $B$ magnitude (solid line). The blue dashed line indicates the expected distribution 
for edge-on spiral galaxies (see the text).
}
\label{fig3}
\end{figure}

Figure\,4 shows the distributions of the sample
galaxies by $h_r$ and $h_z$ values expressed in kpc in the $i_{775}$ filter.
The mean values of these distributions are 
$\left< h_r \right> = 1.97 \pm 0.92$ kpc and 
$\left< h_z \right> = 0.49 \pm 0.16$ kpc. If we restrict
ourselves only to the objects of $eon$ and $fit$ classes 1
and 2 (the number of such galaxies in the sample
is 22), then $\left< h_r  \right> = 2.39 \pm 0.89$ kpc and 
$\left< h_z \right> = 0.52 \pm 0.20$ kpc. The above scales look 
typical for edge-on nearby galaxies (Bizyaev et al. 2014).

The mean ratio of the radial scale lengths in the
$V_{606}$ and $i_{775}$ bands is 1.08$\pm$0.08, implying the 
existence of a color gradient -- the stellar disks of distant
galaxies are, on average, bluer to the periphery. This
feature is typical for the disks of nearby spiral galaxies.
The ratio of the vertical scale lengths in the same filters exhibits 
no noticeable wavelength dependence:
$\left< h_z(V_{606})/h_z(i_{775}) \right> = 0.97 \pm 0.07$. 
This is also consistent with the data for nearby galaxies (see, e.g.,
Bizyaev et al. 2014).

\begin{figure}
\includegraphics[width=0.7\textwidth, angle=-90, clip=]{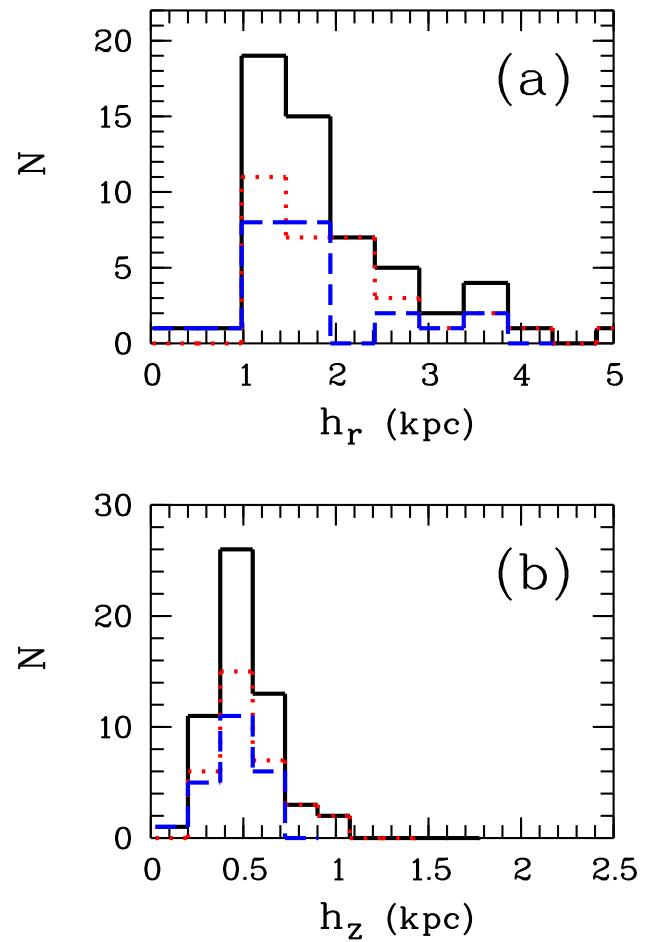}\\
\caption{Distributions of the sample galaxies in (a) radial and (b) 
vertical disk scales (in kpc). The scales are given in the $i_{775}$ filter. 
Different lines correspond to different subsamples of galaxies 
(see the caption to Fig.\,2).}
\label{fig4}
\end{figure}

\subsection{Sample completeness}

Obviously, our sample of edge-on galaxies in the HUDF is incomplete. 
This incompleteness should be most pronounced for faint and poorly resolvable
galaxies, in which the orientation of the stellar disks with respect to 
the line of sight is difficult to determine.

We will use two methods to roughly estimate the expected number of edge-on 
galaxies in the HUDF.

On the one hand, consider the general statistics of galaxies in the HUDF. 
According to Coe et al. (2006), there are $\sim$8000 galaxies in this field. 
The total number of spiral galaxies selected by their spectral energy 
distribution with an apparent F606W magnitude brighter than 27.$^{\text{m}}$5 
(this corresponds to the faintest galaxy in our sample) and $z \leq 2$ is 
1233. Assuming a random orientation of the galactic planes, we
can roughly estimate the relative fraction of edge-on galaxies (with an 
inclination between the line of sight and the normal to the disk plane 
$\geq 85^{\rm o}$) to be $\mid$ cos\,90$^{\rm o}$ -- cos\,85$^{\rm o}\mid$ 
= 0.087. Consequently, the expected number of edge-on spiral galaxies 
in the HUDF is 1233$\times$0.087 $\approx$ 10$^2$.

On the other hand, we can take the luminosity function of nearby edge-on 
spiral galaxies and estimate how many such objects should be observed
toward the HUDF. For our estimation we took the luminosity function of 
spiral galaxies based on data from the 2dF survey (Kroton et al. 2005). 
According to Kroton et al. (2005), the total space density of local spiral 
galaxies in the range of absolute magnitudes from 
$M(B) = -15^{\text{m}}$ to $M (B) = -21^{\text{m}}$
(Fig.\,3) is 0.022 Mpc$^{-3}$. Consequently, the space density of edge-on 
galaxies is 0.022 $\times$ 0.087 = 0.002 Mpc$^{-3}$. Having integrated 
this space density towards the HUDF (its angular size is $\sim 10^{-6}$ sr), 
we found that within $z \leq 2$ about 90 galaxies should be observed in the 
field. The expected distribution of these 90 galaxies in luminosity
is indicated in Fig.\,2 by the blue dashed line.

It can be seen from Fig.\,3 that for bright (with $M(B) \leq -18^{\text{m}}$) 
objects the number of galaxies selected in the HUDF roughly agrees with 
the expected one. The observational selection for fainter galaxies is
apparently much stronger. Consequently, for bright galaxies our sample is 
probably relatively complete, while many galaxies can be missed among 
the fainter objects.

It is worth noting that the above reasoning is not too reliable, because 
in Fig.\,3 we compare the observed luminosities of distant edge-on galaxies 
with the luminosities of nearby galaxies. Because of their edge-on 
orientation, the distant galaxies look fainter by $\sim 1^{\text{m}}$ than 
the face-on galaxies. On the other hand, however, the galaxies at $z \sim 1$ 
should be brighter than the nearby objects approximately by 1$^{\text{m}}$ 
due to the evolution of their luminosity. Both effects can partly
compensate each other out and, therefore, for illustrative
purposes we still compare the luminosities of the nearby and distant 
galaxies in Fig.\,3.

Another factor that is difficult to take into account is the evolution 
of the properties of the spiral galaxies themselves. Because of this 
effect, many of the distant galaxies that look irregular and asymmetric 
at $z \sim 1$ can evolve into typical spiral galaxies with thin stellar
disks by $z \sim 0$.

\subsection{Relative thickness of the stellar disks}

The mean ratio of the radial and vertical exponential disk scales 
for the entire sample (58 galaxies) in the $i_{775}$ band is 
$\left< h_r/h_z \right> = 4.02 \pm 1.28$. If we restrict ourselves only 
to the objects of classes 1 and 2, which characterize the probability of 
assignment to edge-on galaxies and the decomposition quality, then 
$\left< h_r/h_z \right> = 4.61 \pm 0.98$ (22 galaxies).
In the $V_{606}$ filter the corresponding quantities are
$\left< h_r/h_z \right> = 4.46 \pm 1.45$ and 
$\left< h_r/h_z \right> = 5.27 \pm 1.23$.

These mean values look smaller (i.e., the galactic disks are thicker) than 
those for spiral galaxies in the nearby Universe. For example, in the biggest
present-day catalog of edge-on galaxies containing more than 5000 objects 
(Bizyaev et al. 2014), the mean values of this ratio vary from 6.34 in $i$ 
to 7.14 in $g$ (here we took into account the fact that $h_z = z_0/2$; 
$g$ and $i$ are the SDSS\footnote{http://www.sdss.org} filters). Other 
samples of edge-on galaxies also suggest thinner stellar disks of nearby 
spiral galaxies: for example, 
7.4$\pm$2.6 (the $I$ filter; de Grijs 1998), 
16$\pm$4 (Sc/Sd galaxies in the $R$ filter; Schwarzkopf and Dettmar 2000),
7.3$\pm$2.2 ($I$; Kregel et al. 2002), 
9.6 ($K$; Bizyaev and Mitronova 2002), 
7.1 ($J$; Mosenkov et al. 2010),
8.26$\pm$3.44 (De Geyter et al. 2014), and 
8.81$\pm$2.78 (Peters et al. 2017). 
In the last two papers the relative disk thicknesses were obtained by 
simultaneously modeling the galaxies in the SDSS $g$, $r$, $i$, and $z$
filters. Kregel et al. (2002), De Geyter et al. (2014), and Peters et al. 
(2017) used the same photometric model as that in our paper to describe 
the structure of the galaxies. The data from the remaining papers
were recalculated by taking into account the ratio $h_z = z_0/2$.

Note that we compare the observed relative thicknesses of the galaxies 
from the HUDF with those for the nearby edge-on galaxies. This is because 
the stellar disks seen edge-on look more extended due to the
integration of emission along the line of sight. This effect can introduce 
certain systematics into the radial scale lengths measured by different 
methods. For example, Padilla and Strauss (2008) and Rodriguez
and Padilla (2013) estimated the thickness of spiral galaxies by studying 
the distribution of galaxies from SDSS in apparent flattening. As the 
apparent flattening these authors took the SDSS axial ratio found
by fitting the galaxies with an exponential model. According to the first 
and second papers, the true flattening of the disks of spiral galaxies 
is 0.21$\pm$0.02 and 0.27$\pm$0.009, respectively. These values correspond 
to $h_r/h_z = 4.8$ and 3.7 are close to our data for distant galaxies. 
On the other hand, detailed modeling of the structure of nearby 
edge-on galaxies is in conflict with such large stellar disk thicknesses
(see the references above).

Figure\,5 shows the positions of our edge-on galaxies with $eon$ and $fit$ 
classes equal to 1 and 2 on the absolute magnitude -- $h_r/h_z$
plane in the $i_{775}$ band. The same figure displays the data
from the catalog of nearby edge-on galaxies (Bizyaev et al. 2014). 
The galaxies from the HUDF are seen to be located along the lower 
envelope of the distribution for nearby spiral galaxies, i.e., where 
there are the thickest observed disks. Thin stellar disks with an
exponential scales ratio of $\approx$10 are apparently very rare among the 
galaxies at $z \approx 1$.

\begin{figure}
\includegraphics[width=0.35\textwidth, angle=-90, clip=]{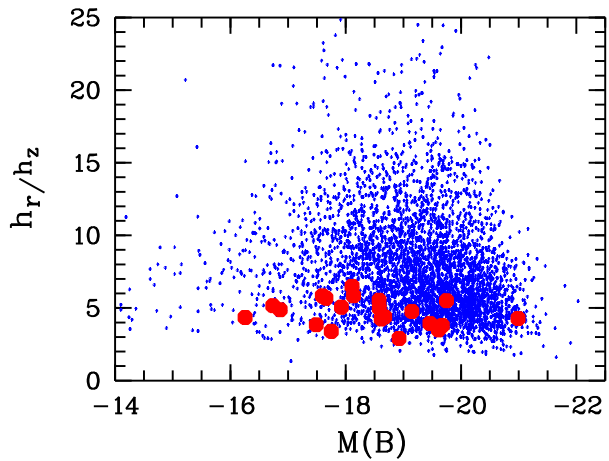}\\
\caption{Distribution of the galaxies from the HUDF on the galaxy absolute 
magnitude $M(B)$ -- radial-to-vertical stellar disk scales ratio plane 
in the $i_{775}$ filter (red circles). The dots indicate the characteristics 
of nearby spiral galaxies in the $g$ band from Bizyaev et al. (2014).}
\label{fig5}
\end{figure}

To a first approximation, the reduced ratio $h_r/h_z$ for distant galaxies 
can be explained by two factors: (1) an increased (in absolute terms) 
thickness of their disks and (2) shorter disks in the radial direction.

Figure\,6 compares the characteristics of the galaxies from the HUDF 
displayed in Fig.\,5 with the parameters of nearby objects on the galaxy 
absolute magnitude -- vertical exponential scale height (in kpc)
and absolute magnitude -- radial scale length (in kpc) planes.

It can be seen from Fig.\,6a that the distant spiral galaxies, though 
with a large scatter, generally follow the luminosity -- stellar disk 
thickness relation for objects in the nearby Universe. For the radial
scale lengths (Fig.\,6b) the situation looks differently:
the characteristics of relatively faint distant galaxies with 
$M(B) \geq -18^{\text{m}}$ lie in the same region as that for
nearby objects, while brighter galaxies exhibit relatively short stellar 
disks.

\begin{figure}
\includegraphics[width=0.7\textwidth, angle=-90, clip=]{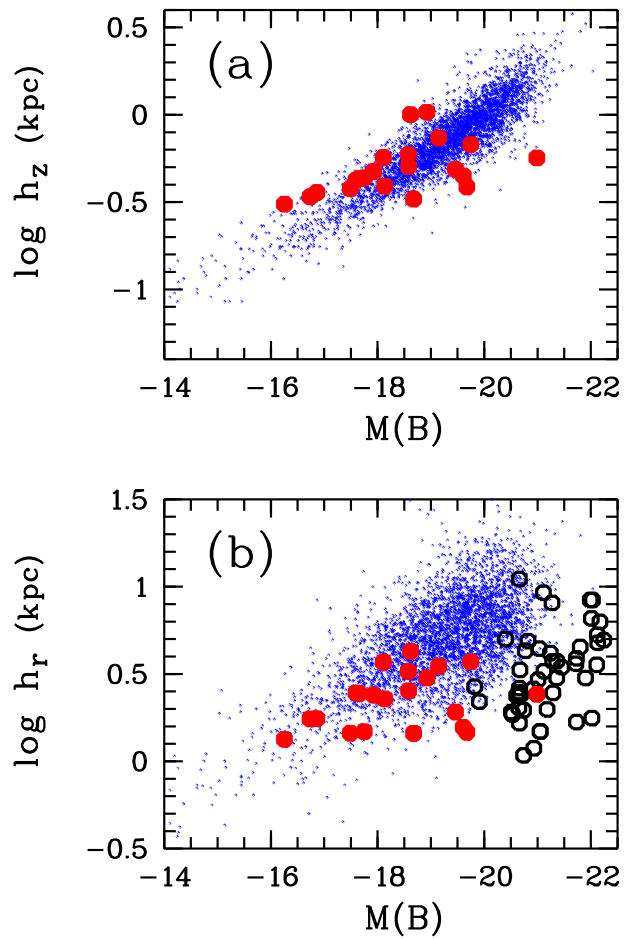}\\
\caption{Distribution of the galaxies from the HUDF (red circles) 
on the (a) $M(B)$ -- disk scale height and (b) $M(B)$ -- disk scale 
length planes. The scales refer to the $i_{775}$ band. The open circles 
indicate the parameters of distant spiral galaxies from Miller et al. (2011) 
($z_{850}$ filter). The dots indicate the characteristics of nearby 
spiral galaxies in the $g$ band from Bizyaev et al. (2014).
}
\label{fig6}
\end{figure}

To check this feature, we plotted the characteristics of 49 spiral 
galaxies at $z = 0.7 - 1.3$ ($\left< z \right> = 0.92 \pm 0.14$) 
from Miller et al. (2011) on the $M(B) - h_r$ plane
(the open circles in Fig.\,6b). The galaxies from Miller et al. (2011) 
have an arbitrary (not edge-on) orientation and they were selected in the 
GOODS field of the Hubble Space Telescope. It can be clearly seen
from the figure that the distant objects from this paper lie on the 
$M(B) - h_r$ plane below the nearby galaxies and form a single sequence 
with the galaxies from the HUDF that deviates from the sequence for 
the objects at $z \sim 0$. The galaxies from Miller et al. (2011) are,
on average, brighter than the objects of our sample (the correction for 
internal absorption does not compensate for the luminosity difference), 
so that, on the whole, the results of our papers complement each
other.

In Figs.\,5 and 6 we compare the observations of distant galaxies in 
the $i_{775}$ filter with the data for nearby galaxies in the $g$ filter 
(the effective wavelength of the filter is $\approx 4600$ \AA).
This is not quite correct since, due to cosmological redshift,
the observed $i_{775}$ filter at $z \approx 1$ corresponds to 
wavelength range $\approx$3500--4000 \AA\, in the galaxy rest frame.
Since the vertical scale height of the stellar  disks depends weakly 
on wavelength (see above), the difference between the color bands plays 
no major role for them. The exponential scale lengths show a
wavelength dependence: the values of $h_r$, on average,  decrease with 
increasing wavelength. Given this effect, the difference between the 
distant and nearby galaxies in Figs.\,5 and 6 is even more pronounced.

Thus, our results in combination with the data from Miller et al. (2011) 
may provide evidence for {\it differential} evolution of the radial sizes 
of spiral galaxies at $z \leq 1$: the low-luminosity objects show no 
evidence of evolution, while the bright spiral galaxies from $z \sim 1$ 
to the present epoch should grow by a factor of 2--3. On the other hand, 
the scale height of spiral galaxies shows no evidence of noticeable 
evolution at $z \leq 1$ (Fig.\,6a). Consequently, the increased relative 
thickness of the stellar disks of spiral galaxies at $z \sim 1$ is 
explained primarily by the smaller radial sizes of their disks.

The results obtained are consistent with the numerical simulations
within $\Lambda$CDM cosmology. For example, Brook et al. (2006) showed 
that at $z \sim 1$ the scale heighs of the stellar disks are already
close to the present-day ones, while the radial scales 
are noticeably shorter. The quantitative agreement between the results 
looks good. For example, according to table\,2 from Brook et al. (2006), 
a spiral galaxy at $z \sim 0.9$ with $M(B) = -21.^{\text{m}}1$ (seen
edge-on, the galaxy will be fainter approximately by 1$^{\text{m}}$), 
$h_r$ = 2.9 kpc, 
and $h_z$ = 0.63 kpc will evolve by $z = 0$ into a galaxy with
$M(B) = -19.^{\text{m}}7$, $h_r$ = 4.1 kpc, and $h_z$ = 0.65 kpc.
Thus, between $z \sim 0.9$ and the current epoch the
relative thickness of the model galaxy changed from
$h_r/h_z$ = 4.6 to 6.3, with this change having occurred
due to the growth of the galaxy in the radial direction.

Based on numerical simulations, Brooks et al. (2011) showed that the 
evolution of disk galaxies depends on their mass. The massive spiral 
galaxies at $z \leq 1$ mostly grow in the radial direction; for the low-mass
ones the change in their sizes is less pronounced, but, on the other hand, 
the luminosity changes more dramatically (see table\,3 in Brooks et al. 
(2011)). The sizes of spiral galaxies increase due to the external
accretion of matter and the merging of satellites.

\section{Conclusions}

Based on an analysis of the HUDF images, we produced a sample of 58 
candidates for edge-on spiral galaxies at a mean redshift $z \sim 1$. 
For all galaxies we analyzed the 2D brightness distributions in the
$V_{606}$ and $i_{775}$ filters and determined the radial and
vertical exponential scales of the brightness distribution.

Our main results are as follows:

-- The scale height of the stellar disks of spiral galaxies shows no 
significant evolution at $z \leq 1$.

-- The relative thickness of the disks of distant galaxies, on average, 
exceeds the relative thickness of the disks of nearby spiral galaxies. 
Thin stellar disks at $z \sim 1$ are apparently very rare.

-- We obtained evidence for differential evolution of the exponential 
scale lengths of the stellar disks of galaxies: the bright spiral 
galaxies at $z \sim 1$ look shortened compared to the nearby objects; 
the low-luminosity galaxies show no evidence of evolution.

The results of this paper are based on a small sample of galaxies and, 
undoubtedly, need to be confirmed with a larger number of objects. 
On the whole, our observational results are consistent with the current
views of the evolution of the disk subsystems of galaxies and they can 
be used to test various models for the evolution of spiral galaxies.

\vspace{0.5cm}

This work was supported by the Russian Science Foundation 
(grant no. 19-12-00145).

We are grateful to A.V. Mosenkov for useful comments.

\vspace{0.5cm}

\section*{REFERENCES}

\scriptsize

\noindent
1. S.V.W. Beckwith, M. Stiavelli, A.M. Koekemoer,
J.A.R. Caldwell, H.C. Ferguson, R. Hook, R.A. Lucas, L.E. Bergeron, 
et al., Astron. J. 132, 1729 (2006). \\

\noindent
2. E. Bertin and S. Arnouts, Astron. Astrophys. Suppl.
Ser. 117, 393 (1996). \\

\noindent
3. J. Bicker, U. Fritze-v.Alvensleben, C.S. Moller, and
K. J. Fricke, Astron. Astrophys. 413, 37 (2004). \\

\noindent
4. D. Bizyaev and S. Mitronova, Astron. Astrophys. 389,
795 (2002). \\

\noindent
5. D.V. Bizyaev, S.J. Kautsch, A.V. Mosenkov,
V.P. Reshetnikov, N.Ya. Sotnikova, N.V. Yablokova,
and R. W. Hillyer, Astrophys. J. 787, 24 (2014).\\

\noindent
6. D.V. Bizyaev, S.J. Kautsch, N.Ya. Sotnikova,
V.P. Reshetnikov, and A.V. Mosenkov, Mon. Not. R.
Astron. Soc. 465, 3784 (2017).\\

\noindent
7. Ch.B. Brook, D. Kawata, H. Martel, B.K. Gibson,
and J. Bailin, Astrophys. J. 639, 126 (2006).\\

\noindent
8. A.M. Brooks, A.R. Solomon, F. Governato, J. McCleary, 
L.A. MacArthur, C.B.A. Brook, P. Jonsson, T.R. Quinn, et al., 
Astrophys. J. 728, 51 (2011).\\

\noindent
9. D. Coe, N. Benitez, S.F. Sanchez, M. Jee,
R. Bouwens, and H. Ford, Astron. J. 132, 926 (2006).\\

\noindent
10. D.M. Elmegreen, B.G. Elmegreen, D.S. Rubin, and
M.A. Schaffer, Astrophys. J. 631, 85 (2005).\\

\noindent
11. B.G. Elmegreen and D.M. Elmegreen, Astrophys.
J. 650, 644 (2006).\\

\noindent
12. P. Erwin, Astrophys. J. 799, 226 (2015).\\

\noindent
13. G. de Geyter, M. Baes, P. Camps, J. Fritz, I. de Looze,
Th.M. Hughes, S. Viaene, and G. Gentile, Mon. Not.
R. Astron. Soc. 441, 869 (2014).\\

\noindent
14. R. de Grijs, Mon. Not. R. Astron. Soc. 299, 595
(1998).\\

\noindent
15. H. Inami, R. Bacon, J. Brinchmann, J. Richard,
T. Contini, S. Conseil, S. Hamer, M. Akhlaghi, et al.,
Astron. Astrophys. 608, A2 (2017).\\

\noindent
16. M. Kregel, P.C. van der Kruit, and R. de Grijs, Mon.
Not. R. Astron. Soc. 334, 646 (2002).\\

\noindent
17. J.E. Krist, R.N. Hook, and F. Stoehr, Proc. SPIE
8127, 81270J (2011).\\

\noindent
18. P.C. van der Kriut, and L. Searl, Astron. Astrophys.
95, 105 (1981).\\

\noindent
19. D.I. Makarov, N.A. Zaitseva, and D.V. Bizyaev,
Mon. Not. R. Astron. Soc. 479, 3373 (2018).\\

\noindent
20. S.H. Miller, K. Bundy, M. Sullivan, R.S. Ellis, and
T. Treu, Astrophys. J. 741, 115 (2011).\\

\noindent
21. A.V. Mosenkov, N.Ya. Sotnikova, and V.P. Reshetnikov, 
Mon. Not. R. Astron. Soc. 401, 559 (2010).\\

\noindent
22. A.V. Mosenkov, F. Allaert, M. Baes, S. Bianchi,
P. Camps, G. de Geyter, I. de Looze, J. Fritz, et al.,
Astron. Astrophys. 592, A71 (2016).\\

\noindent
23. P.C. Peters, G. de Geyter, P.C. van der Kruit, and
K.C. Freeman, Mon. Not. R. Astron. Soc. 464, 48
(2017).\\

\noindent
24. M. Rafelski, H.I. Teplitz, J.P. Gardner, D. Coe,
N.A. Bond, A.M. Koekemoer, N. Grogin, P. Kurczynski, et al., 
Astron. J. 150, 31 (2015).\\

\noindent
25. V.P. Reshetnikov, R.-J. Dettmar, and F. Combes,
Astron. Astrophys. 399, 879 (2003).\\

\noindent
26. U. Schwarzkopf and R.-J. Dettmar, Astron. Astrophys. 361, 451 (2000).\\

\noindent
27. M. Sirianni, M.J. Jee, N. Benitez, J.P. Blakeslee,
A.R. Martel, G. Meurer, M. Clampin, G. de Marchi,
et al., Publ. Astron. Soc. Pacif. 117, 1049 (2005).\\

\noindent
28. R.B. Tully, M.J. Pierce, J.-Sh. Huang, W. Saun-
ders, M.A.W. Verheijen, and P.L. Witchalls, Astron.
J. 115, 2264 (1998).\\

\noindent
29. A.V. Zasov, D.I. Makarov, and E.A. Mikhailova,
Astron. Lett. 17, 374 (1991).

\end{document}